\documentstyle[fleqn,12pt]{article}
\title{Closed Strings with Low Harmonics and Kinks}
\author{R.W.Brown, M.E.Convery, S.A.Hotes, M.G.Knepley, and L.S.Petropoulos}
\date{}
\begin{document}
\maketitle
{\large\centerline{\it Department of Physics, Case Western
  Reserve University,}
\centerline{\it Cleveland, Ohio 44106}}
\bigskip\bigskip
\begin{abstract}
   Low-harmonic formulas for closed relativistic strings are given.
General parametrizations are
presented for the addition of second- and third-harmonic waves to
the fundamental wave. The method of determination of the
parametrizations is based upon a product representation found for
the finite Fourier series of string motion in which the
constraints are automatically satisfied.  The construction of
strings with kinks is
discussed, including examples.  A procedure is laid out for the
representation of kinks that arise from self-intersection, and subsequent
intercommutation, for harmonically parametrized cosmic strings.
\end{abstract}
\vfill
\newpage
\section{Introduction}
\label{sec:Intro}
	The relativistic string model has been at the heart of much
of theoretical elementary particle physics for the past decade,
both as a model of elementary particles and as a description of
cosmic string defects postulated to have been produced in the
early universe.  The cosmic string hypothesis~\cite{Ki} in
particular continues to attract interest in attempts to
understand the large-scale structure observed in galaxy
distributions.  Recently it has been argued that cosmic strings are
consistent with the non-uniformities
observed in the background radiation~\cite{COBE}.  In addition,
the string model has proven to be a well-spring of discoveries in
connecting different mathematics to physics, and as a vehicle for
exploring new mathematics and new mathematical techniques.

	In the study of closed cosmic strings, one subject of
interest has been the harmonic solutions to the relativistic string
equations in flat space.  Whenever there are damping mechanisms at work,
one expects that
higher harmonics would get relatively suppressed.  One can then
consider a finite Fourier
series - a series of a finite number of harmonics - and derive the
Fourier coefficients required by the constraint equations in a given
gauge.  Although an infinite number of
harmonics is considered at the outset in the study of a quantized
string - in order to have a complete basis for distributions and to
preserve locality - certain fundamental string issues may be conveniently
studied with strings containing a finite number of harmonics.

	There is a specific mechanism that regularly dumps power into
the higher harmonics of cosmic strings, in the midst of the damping.
As a result of the intercommutation of intersecting
strings, infinite and closed, scars develop in the form of distinct
kinks~\cite{Sh,Ma,LM}.  The kinks present in generic
cosmic loops at early stages of the universe do eventually decay away,
not by radiation alone~\cite{GV}, but by back reaction to that radiation
{}~\cite{Qu}.

 	In the course of cosmic string studies concerning
radiation, self-intersection, and
black-hole formation, it has proven useful to construct loop
solutions for a few low
harmonics~\cite{KT,T,Bu,VV} . A systematic
investigation of more and more general parametrizations of the
low-lying harmonics has been undertaken by our group over the past
few years~\cite{CDH,DES,Br}. This  has led to a general
solution for any closed string with an arbitrarily large number $N$
for the largest harmonic to be
included~\cite{BD,BCD,BRT}.  The result corresponds to
finding the most general Fourier series of a unit vector, given
an arbitrary but finite number of harmonics.
Consistent with what we have come to
expect from string theory, we have a new mathematics tool useful for other
applications.

	In this paper we put the new methodology to work.  We
construct general $N\leq 3$ solutions for closed loops.  A
catalog of previous
solutions is presented in terms of the product
representation parameters, and
the inclusion of kinks is considered (see Ref. \cite{Sch,All} for earlier
discussions).
A construction algorithm for string solutions
containing a single left or right moving
kink is illustrated.
We analyze the fact that when kinks are created through
intercommutation, the kink will split into right and left traveling
pieces.  A general procedure is provided for analytically describing
these resulting equations of motion, and an example is given.

	The flat-space string equations and their Fourier
analysis comprise Sections II and III, respectively. Section IV
concentrates upon the exact solutions for the general case of
strings with first, second, and third harmonics using the special
rotation form described in Section III. Strings previously
developed by others are rewritten within the framework of the
rotation form in Section V, and shown to be a subset of this more
general procedure.  Figures  are used to illustrate loop motion for various
parametrizations.  We next introduce the concept of a kinked
string.  In section VI we show how single kink
strings may be introduced by fine-tuning the parametrization of
the right (or left) traveling wave as it traverses the Kibble-Turok
sphere.  Section VII describes the process of intercommutation and
an example, employing
the results of the previous section.
Section VIII contains some concluding remarks.

\vskip5ex

\section{Closed String Review}
\label{sec:StringEq}

In an
orthonormal gauge~\cite{GGRT}, the string position $\vec r (\sigma
,t)$ satisfies the wave equation $\vec r_{tt} - \vec r_{\sigma
\sigma} = 0$\ ($f_t \equiv \partial f/ \partial t$, etc.), and
the constraints are transverse motion $\vec r_t \cdot \vec
r_\sigma =0$ and unit energy density $r_t^2 + r_\sigma^2 =1$.
Within this Lorentz frame, the string parameter $\sigma$ has the
same units as the time $t$, and we scale to the interval, $0
\leq \sigma \leq 2\pi$.  The general
``right-going" ($u=\sigma-t$) plus ``left-going" ($v=\sigma+t$)
wave solution is \begin{equation}\vec r={1\over 2} [\vec
a(u)+\vec b(v)], \label{eq:gen_soln}\end{equation} in what has
become rather standard notation.  The constraints imply
\begin{equation} \vec{a}^{\prime \; 2}=\vec{b}^{\prime \; 2}=1;
\label{eq:unit}\end{equation}
that is, $\vec a^{\,\prime}$ and $\vec
b^{\,\prime}$ traverse the Kibble-Turok sphere~\cite{KT}.  We see
that we must have both left- and right-going waves, and they
must be ``equally
weighted" in the sense that both derivatives have unit magnitude.

The overall spatial periodicity of a closed loop of string, $\vec
r(\sigma +  2\pi,t) = \vec r(\sigma,t)$ or
\begin{equation}\vec a(u+2\pi)+ \vec b(v+2\pi) = \vec a(u)+
  \vec b(v),
\label{eq:per}\end{equation}
holds true up to linear (c.m.) terms for
$\vec a$ and $\vec b$ as well.  Consider two values of $u$, but with
$v$ fixed.  Eq.(3) yields
\begin{equation}\vec a(u_1+2\pi)-\vec a(u_2+2\pi)=
  \vec a(u_1)-\vec a(u_2),
\label{eq:aper}\end{equation}
and,
\begin{equation}\vec a^{\,\prime}(u+2\pi)=\vec a^{\,\prime}(u).
\label{eq:alim}\end{equation}
Similarly,
\begin{equation}\vec b^{\,\prime}(v+2\pi)=\vec b^{\,\prime}(v).
\label{eq:blim}\end{equation}
That is, the unit vectors of eq.(2.2) are
periodic.  From eqs.(3), (5), and (6),
\begin{equation}\vec a(u+2\pi)=\vec a(u)+\vec k,
\label{eq:aconst}\end{equation}
\begin{equation}\vec b(v+2\pi)=\vec b(v)-\vec k.
\label{eq:bconst}\end{equation}
We can write
\begin{equation}\vec a(u)\equiv\vec a_p(u)+\vec k u/2\pi,
\label{eq:ap}\end{equation}
\begin{equation}\vec b(v)\equiv\vec b_p(v)-\vec k v/2\pi,
\label{eq:bp}\end{equation}
where $\vec a_p$ and $\vec b_p$ are periodic functions of their
arguments, with period $2\pi$.  In a Fourier series, $\vec
a_p^{\,\prime}$ and $\vec b_p^{\,\prime}$ have no zero harmonic
terms.  The derivatives,
\begin{equation}\vec a^{\,\prime}=\vec a_p^{\,\prime}+\vec k/2\pi,
\label{eq:apprime}\end{equation}
\begin{equation}\vec b^{\,\prime}=\vec b_p^{\,\prime}-\vec k/2\pi,
\label{eq:bpprime}\end{equation}
are indeed periodic,
and $\pm\vec k/2\pi$ are identified as the respective zero harmonics.

Finally, we have
\begin{equation}\vec r(\sigma,t)=\vec r_p(\sigma,t)+\vec Kt,
\label{eq:rp}\end{equation}
with
\begin{equation}\vec r_p \equiv {1 \over 2} [\vec a_p(u)+
  \vec b_p(v)],
\label{eq:gen_solnrp}\end{equation}
\begin{equation}\vec K \equiv -\vec k/\pi.
\label{eq:K}\end{equation}
Eq.(13) is the generic form for closed loops: periodic
left-going and right-going superimposed on uniform (c.m.) motion.

Now we address temporal periodicity.  Modulo the c.m. motion, the
string certainly repeats in $2\pi$ time intervals.
{}From (4),
\begin{equation}\vec r_p(\sigma,t+2\pi)=\vec r_p(\sigma,t).
\label{eq:rpper}\end{equation}
But in fact the effective time period is half of this~\cite{KT} because
under $\sigma \rightarrow \sigma + \pi$, $t \rightarrow t + \pi$,
we have $u \rightarrow u$, $v \rightarrow v + 2\pi$,
\begin{equation}\vec r_p(\sigma+\pi,t+\pi)=\vec r_p(\sigma,t).
\label{eq:rpper2}\end{equation}
The string ``looks" exactly the same every time interval $T$,
\begin{equation}T=\pi.
\label{eq:T}\end{equation}
The first half of the string $(0<\sigma<2\pi)$ switches places with the
second half $( \pi < \sigma<2\pi)$ every time $T$.
The individual unit vectors eqs.(9) and (10), however,
have period $2\pi$.

\vskip5ex

\section{Fourier Analysis and Product Representation}
\label{sec:Fourier}

We are thus led to consider the Fourier analysis of $\vec
a^{\,\prime}$ and $\vec b^{\,\prime}$, the integration of which
will give us the string configuration.  We restrict ourselves to
a finite number of harmonics according to the discussion in the
introduction. The problem of finding the harmonic coefficients in
the finite Fourier series for a periodic vector whose magnitude
is fixed has just recently been solved.  The product
representation of Ref.~\cite{BD} automatically satisfies the
magnitude constraint, exhibits the correct degrees of freedom,
and gives the general solution.

Consider a periodic unit vector $\hat u_N(s)=\hat u_N(s+2\pi)$
defined by the $N$-harmonic real series,
\begin{equation} {\hat u}_N(s) =\vec Z + \sum_{n=1}^N
  (\vec A_n \cos{ns} + \vec B_n \sin{ns}),
\label{eq:Fourier}\end{equation}
with $N$ arbitrary.  The constraint $(\hat u_N)^2=1$ requires a set
of nonlinear relations among the vector coefficients
in eq.(19).  In terms of the real basis, we have the
$4N+1$ real equations
\begin{equation}\sum_{n=m-N}^N (\vec \alpha_n \cdot \vec \alpha_{m-n}
  -\vec \beta_n \cdot \vec \beta_{m-n})=4\delta_{m0},
   \quad m=0,1,\dots,2N,
\label{eq:FourCond1}\end{equation}
\begin{equation}\sum_{n=m-N}^N (\vec \alpha_n \cdot \vec \beta_{m-n}
  +\vec \beta_n \cdot \vec \alpha_{m-n})=0,
   \quad m=1,\dots,2N.
\label{FourCond2}\end{equation}
Here,
\begin{equation}\vec \alpha_n = \vec \alpha_{-n} = \vec A_n,
  \quad \vec \beta_n = -\vec \beta_{-n} = \vec B_n,
  \qquad  n \not= 0,
\label{eq:AB}\end{equation}
\begin{equation}\vec \alpha_{0} = 2 \vec Z, \quad \vec \beta_{0} =0.
\label{eq:Z}\end{equation}

The product representation which solves eq.(19) can be
written in terms of standard matrices such as $R_\xi(\psi)$, which
refers to a rotation of a vector through the angle $\psi$ about the
$\xi$-axis (or of the coordinate system through $-\psi$).  Making a
specific choice of a constant unit vector which we rotate in
succession, we have
\begin{equation}\hat u_N(s)=r_N R_z(s) \rho_N R_z(s) \rho_{N-1}
  \dots R_z(s) \rho_1 \hat z.
\label{eq:prod}\end{equation}
Here,
\begin{equation}\rho_i=R_z(-\theta_i) R_x(\phi_i) R_z(\theta_i).
\label{rho}\end{equation}
We can let $\theta_1=\pi/2$, or
\begin{equation}\rho_1=R_y(-\phi_1).
\label{eq:rho1}\end{equation}
Also, there is the overall orientation freedom
\begin{equation} r_N=\left\{ \begin{array}{ll} R_z(\alpha) R_x(\beta)
  R_z(\gamma), & N \geq 1,\\
  R_z(\alpha) R_x(\beta), & N=0.
  \end{array}\right.
\label{eq:rn}\end{equation}
where $\alpha$, $\beta$, and $\gamma$ are additional constants (angle
parameters).

Let us consider how eq.(24) gives the desired properties.
The rotations preserve the
vector magnitude, satisfying the original constraint, and the $N$ factors
of $R_z(s)$ generate $N$ harmonics. In general, this is a
complete and independent representation, although one needs to look
at the detailed proof~\cite{BD,BCD} to see this.  The $2N+2$ independent
degrees of freedom [$(6N+3)-(4N+1)$] are unrestricted angles,
$\theta_i$ and $\phi_i$, whose ranges are independent of each other
($0 \leq \theta_i \leq \pi$, $0 \leq \phi_i \leq 2\pi$). (In
examples, there may exist reflection symmetries, such as $\phi_i
\rightarrow -\phi_i$ or $\pi-\phi_i$ for some of the angles, reducing
the overall range accordingly.) Because the signs can be a source
of confusion in the derivation of the results in the next section, we
list the matrix conventions for active rotations
\begin{equation}\!\!\!\!\!\!\!\!\!\!\! R_x(\theta) = \left(
  \begin{array}{ccc}
  1 & 0 & 0 \\ 0 & c\theta & -s\theta \\ 0 & s\theta & c\theta \\
  \end{array} \right),
  R_y(\theta) = \left( \begin{array}{ccc} c\theta & 0 & s\theta \\
  0 & 1 & 0 \\ -s\theta & 0 & c\theta \end{array} \right),
  R_z(\theta) = \left( \begin{array}{ccc} c\theta & -s\theta & 0 \\
  s\theta & c\theta & 0 \\ 0 & 0 & 1 \end{array} \right),
\end{equation}
with $s\theta \equiv \sin{\theta}$, $c\theta \equiv \cos{\theta}$.

To complete the groundwork for a procedure for constructing closed
strings for a given $N$ (zeroth plus first $N$ harmonics), we note
that the overall rotation $r_N$ can be omitted in the product
representation (3.6) of the unit vector $\vec a$,
\begin{equation}\vec a^{\,\prime}_N(u,\theta_i,\phi_i)=
  \prod_1^N[R_z(u)\rho_i] \hat z.
\label{eq:aprimeprod}\end{equation}
Here, $f(\theta_i)\equiv f(\{\theta_i\})$, etc.  In place of
$r_N$, we consider the $\{\hat x, \hat y, \hat z\}$ basis as
arbitrarily oriented. To obtain final polynomial expressions in
$\sin{\theta_i}$, $\cos{\theta_i}$, etc., it is convenient to use
the iterative property
\begin{equation}\vec a^{\,\prime}_N=R_z(u-\theta_N) R_x(\phi_N)
  R_z(\theta_N) \vec a^{\,\prime}_{N-1}.
\label{eq:aprimeproditer}\end{equation}
Afterwards, it is necessary to separate out the zero harmonic
piece in eq.(11), defined to be $\vec \alpha_N$,
\begin{equation}\vec a^{\,\prime}_N(u,\theta_i,\phi_i)= \vec
  a^{\,\prime}_{p,N} (u,\theta_i,\phi_i) +\vec \alpha_N(\theta_i,
  \phi_i).
\label{eq:aprodp}\end{equation}
We can immediately write down the other unit vector and its zero
harmonic component $\beta_N$
\begin{equation} \begin{array}{rl} \vec b^{\,\prime}
  _N(v,\theta_i^\prime, \phi_i^\prime) & =\vec b^{\,\prime}
  _{p,N}(v,\theta_i^\prime,\phi_i^\prime)
  +\vec \beta_N(\theta_i^\prime,\phi_i^\prime)\\
  \ & =\vec a^{\,\prime}_N(v,\theta_i^\prime,\phi_i^\prime)
  |_{\hat x \hat y \hat z \rightarrow \hat x^\prime \hat y^\prime
  \hat z^\prime}. \\ \end{array} \label{eq:bprodp}
\end{equation}
That is, $u \rightarrow v$, and prime everything else.  Finally,
we must have the periodicity condition (15),
\begin{equation} \vec \alpha_N(\theta_i,\phi_i)=-\beta
  _N(\theta_i^\prime, \phi_i^\prime)= -{1 \over 2} \vec K_N.
\label{eq:zero}\end{equation}
This last condition can be met in two ways.  The first we call the
traveling string case where all $\theta'_i$, $\phi'_i$, $\hat x'$,
$\hat y'$, $\hat z'$ are found so that (33) is satisfied.
The second we refer to as the c.m. frame, where $\vec K_N=0$.  We
find all $\theta_i$, $\phi_i$ such that
\begin{equation}\vec \alpha_N=0,
\label{eq:alpha0}\end{equation}
and $\theta'_i$, $\phi'_i$ such that
\begin{equation}\vec \beta_N=0.
\label{eq:beta0}\end{equation}

\vskip5ex
\section{General Parametrizations of N=0,1,2,3 Closed Strings}
\label{sec:Params}

We begin with a zeroth harmonic and then add
in sequence a first, second, and finally a third harmonic.  In each case
we describe two
different solutions corresponding to moving strings
 and c.m. strings,
as described in the previous section.

\subsection{N=0}
\label{subsec:N=0}

The N=0 trivial zero-harmonic case for eq.(30) is
included if only to highlight the freedom to choose an overall
z-axis direction.  We have $\vec a^{\,\prime}=-\hat z$, $\vec
b^{\,\prime}=\hat z'$, with the periodicity of $\vec r$ in
$\sigma$ forcing $\hat z^{\,\prime}=\hat z$ [recall
eq.(17)].  The result is a point moving at the speed of
light,
\begin{equation}\vec r_0=\hat zt.
\label{eq:N=0}\end{equation}

\subsection{N=1}
\label{subsec:N=1}

The general combination of zeroth plus first harmonic is the N=1
case in eq.(3.12). In simpler notation ($\phi_1
\equiv~-\theta , -\theta'$ for $\vec a^{\,\prime}, \vec
b^{\,\prime}$, respectively), matrix multiplication leads to the intermediate
answers
\begin{equation}\vec a^{\,\prime}=\sin{\theta}\cos{u}\,\hat x
  + \sin{\theta}\sin{u}\,\hat y +\cos{\theta}\,\hat z,
\label{eq:aprime1}\end{equation}
\begin{equation}\vec b^{\,\prime}=\sin{\theta'}\cos{v}\,
  \hat x^{\,\prime} \pm\sin{\theta'}\sin{v}\,
  \hat y^{\,\prime}+\cos{\theta'}\,\hat z^{\,\prime}.
\label{eq:bprime1}\end{equation}
Besides the $z$ freedom, we have chosen a particular handedness
for the time rotation of $\vec a^{\,\prime}$ for a given
$\theta$.  But then $\vec b^{\,\prime}$ can have both right- and left-handed
circulation.

\vskip2ex
\subsubsection{Traveling string}
\label{subsubsec:N=1Trav}
For the ``moving string" procedure of satisfying $\sigma$ periodicity, $\hat
z^{\,\prime}=\hat z$ and $\theta'=\theta+\pi$.  Integration gives
the traveling loop solution for both circulations,
\begin{equation}\vec r^{\,\pm}={1 \over 2}\sin{\theta}
  [(\sin{u}-\sin{v})\hat x+(-\cos{u}\pm\cos{v})\hat y]
  -\cos{\theta}\hat zt.
\label{eq:N=1Trav}\end{equation}
(In such deliberations, any rotation by an angle $\chi$ of $\hat
x^{\,\prime} -\hat y^{\,\prime}$ relative to $\hat x-\hat y$ can
be absorbed into redefinitions (shifts) of $\sigma$ and $t$:
$\hat x^{\,\prime}\cos{v}+\hat y^{\,\prime}\sin{v}=\hat x
\cos{(v+\chi)}+\hat y\sin{(v+\chi)}\rightarrow\hat x\cos{v}+\hat
y\sin{v}$, for $\sigma\rightarrow   \sigma-\chi/2$, $t\rightarrow
t-\chi /2$.)  Eq.(39) refers to two simple planar
strings with constant c.m. motion that is perpendicular to the
planes.  Rewriting, we see these are, respectively, circles with oscillating
radii,
\begin{equation}\vec r^+=-\sin{\theta}\sin{t}\hat \rho
  (\sigma)-\cos{\theta} \hat zt,
\label{eq:r+}\end{equation}
and uniformly rotating sticks of fixed length,
\begin{equation}\vec r^-=-\sin{\theta}\cos{\sigma}\hat
  \phi (-t)-\cos{\theta} \hat zt.
\label{eq:r-}\end{equation}
Here we have used the cylindrical unit vectors
\[ \hat \rho (A)=\hat x\cos{A}+\hat y\sin{A},\]
\begin{equation} \hat \phi (A)=-\hat x\sin{A}+\hat y\cos{A}.
\label{eq:cyl}\end{equation}
The speed of light is reached at the ends of the sticks or when the
circles pass through zero radius.  The limits $\theta=0,\pi$ give
the trivial case presented in eq.(36).

\vskip2ex
\subsubsection{c.m. string}
\label{subsubsec:N=1CM}

In the ``c.m." procedure, we eliminate the zero harmonics in
eq.(37) and (38) separately and hence the
c.m. motion, by the requirement that $\theta=\theta'=\pi /2$.
Visualizing the most general intersection of two great circles on
the Kibble-Turok sphere, we let $\hat y^{\,\prime}=\hat y$ and
$\hat x^{\,\prime}= \hat x\cos{\psi}+\hat z\sin{\psi}$:
\begin{equation} \vec r^{\,\pm}={1\over 2}[(\sin{u}+\sin{v}
  \cos{\psi})\hat x-(\cos{u}\pm\cos{v})\hat y+\sin{v}
  \sin{\psi}\hat z].
\label{eq:N=1CM}\end{equation}
These strings look like rotating ellipses that oscillate in size,
collapsing periodically to sticks.  They are natural
interpolations between the circle and stick results in
eq.(40), (41), forms to which they reduce in the
$\psi=0,\pi$ limits but without any overall motion.  See Fig. 1.
for an illustration.  It is interesting, however, that we cannot get the
general
c.m. solution by simple limits on the traveling string solution.

\subsection{N=2}
\label{subsec:N=2}

The combination of the three harmonics - zeroth, first, second - illustrates
an interesting point.  In the c.m. string
limit where one eliminates the zeroth
harmonic,
it is seen that no solution is possible for that particular
``half" of the string ($\vec a$ or $\vec b$).  A first harmonic and a
second harmonic cannot coexist.  In fact, the only pair of nonzero harmonics
that can coexist must be in the ratio of three to one~\cite{DES}.

\subsubsection{Traveling string}
\label{subsubsec:N=2Trav}

For a nonzero zero-harmonic contribution, all three harmonics coexist.
In the ``moving string" procedure of dealing with the zero harmonics of
$\vec a^{\,\prime}$ and $\vec b^{\,\prime}$, the $\sigma$ terms
cancel when the angles are related as follows
\begin{equation} \phi'_2=\pm\phi_2, \ \phi'_1=\phi_1+\pi,
  \ \theta'_2=\theta_2, \label{eq:N=2TravCond}
\end{equation}
where the unprimed and primed angles are parameters of $\vec a$
and $\vec b$, respectively.  This gives four strings due to the
two possible values of $\phi_2$  (represented by $\pm_2$) and to
the freedom for $\vec b$ to rotate in a right- or left-handed
direction (represented by $\pm$):
$$ \vec r=\:\left( \begin{array}{c}
  \cos^2{\phi_2\over 2}\sin{\phi_1} \\ 0 \\ 0 \end{array} \right)
  {1\over 4}(\sin{2u}-\sin{2v})
  +\left( \begin{array}{c}
  0 \\ \cos^2{\phi_2\over 2}\sin{\phi_1} \\ 0 \end{array} \right)
  {1\over 4}(-\cos{2u}\pm\cos{2v})$$
\[ \!\!\!\!\!\!+\left( \begin{array}{c}
  -\sin{\phi_2}\cos{\phi_1}\sin{\theta_2} \\
  -\sin{\phi_2}\cos{\phi_1}\cos{\theta_2} \\
   \sin{\phi_2}\sin{\phi_1}\sin{\theta_2} \end{array} \right)
  {1\over 2}(\sin{u}-(\pm_2)\sin{v})\]
\[ \!\!\!\!\!\!+\left( \begin{array}{c}
   \sin{\phi_2}\cos{\phi_1}\cos{\theta_2} \\
  -\sin{\phi_2}\cos{\phi_1}\sin{\theta_2} \\
   \sin{\phi_2}\sin{\phi_1}\cos{\theta_2} \end{array} \right)
  {1\over 2}(-\cos{u}\pm(\pm_2)\cos{v})\]
\begin{equation} \!\!\!\!\!\!+\left( \begin{array}{c}
  -\sin^2{\phi_2\over 2}\sin{\phi_1}\cos{2\theta_2} \\
   \sin^2{\phi_2\over 2}\sin{\phi_1}\sin{2\theta_2} \\
  -\cos{\phi_2}\cos{\phi_1} \end{array} \right) t \,\, .
\end{equation}

\vskip2ex
\subsubsection{c.m. string}
\label{subsubsec:N=2CM}

There are two ways to set both the zero harmonics equal to zero.
\begin{equation} \phi'_2=\phi_2={\pi\over 2},{3\pi\over 2},
  \ \phi'_1=\phi_1=0,\pi \, .
\label{eq:N=2CMCond1}\end{equation}
This reduces the string to a two-parameter $N=1$ string where
$\theta_1=\theta_2$, $\phi_1={\pi\over 2}$,
$\theta_1'=\theta_2'$, $\phi_1'={\pi\over 2}$.

The second set of angles
\begin{equation} \phi'_2=\phi_2=0, \ \phi'_1=\phi_1
  ={\pi\over 2},{3\pi\over 2} \, ,
\label{eq:N=2CMCond2}\end{equation}
give a string with only a second harmonic
\begin{equation} \vec r=\pm{1\over 4} \left( \begin{array}{c}
   (\sin{2u}+\cos{\psi}\sin{2v}) \\ -\cos{2u}\mp\cos{2v} \\
   \sin{\psi}\sin{2v}\\ \end{array} \right),
\label{eq:N=2CM}\end{equation}
which is simply a z-rotation of $u$ on eq.(43).

Other solutions may be obtained by satisfing one of the above
constraints on one half of the string, and another set of
constraints on the other half. For example, setting
$\phi_2 = 0,\ \phi_1 ={\pi\over2},\ \phi'_2 ={\pi\over2},\
\phi'_1 = 0$, we obtain the following string,
\begin{equation} \vec r={1\over4} \left( \begin{array}{c}
  -2(\cos{\theta_2} \cos{u} + \sin{\theta_2} \sin{u}) +
  \cos{\psi} \sin{2v} \\
  2 (\cos{u} \sin{\theta_2} - \cos{\theta_2} \sin{u}) - \cos{2v} \\
  \sin{\psi} \sin{2v} \\
  \end{array} \right).
\label{eq:N=2CM2}\end{equation}

\subsection{N=3}
\label{subsec:N=3}
\subsubsection{Traveling string}
\label{subsubsec:N=3Trav}

The general equations fixing the zero harmonics so that they
cancel are highly nonlinear, and the solution set is very
difficult to define, but the following four solutions are fairly
easy to see:
$$\begin{array}{lllll} \phi'_3=\phi_3, & \theta'_3=\theta_3,
  & \phi'_2=\phi_2, & \theta'_2=\theta_2, & \phi'_1=\phi_1+\pi, \\
  \phi'_3=\phi_3, & \theta'_3=\theta_3, & \phi'_2=-\phi_2,
  & \theta'_2=\theta_2+\pi, & \phi'_1=\phi_1+\pi, \\
  \phi'_3=-\phi_3, & \theta'_3=\theta_3+\pi, & \phi'_2=\phi_2,
  & \theta'_2=\theta_2, & \phi'_1=\phi_1+\pi,  \\
  \phi'_3=-\phi_3, & \theta'_3=\theta_3+\pi, & \phi'_2=-\phi_2,
  & \theta'_2=\theta_2+\pi, & \phi'_1=\phi_1+\pi.
\label{eq:N=3TravCond}\end{array}$$

These all give the same two string equations (in the more concise
form of $\hat \phi$, $\hat \rho$ from (\ref{eq:cyl})),
\[\!\!\!\!\!\!\!\!\!\!\vec r=\quad -\;\frac{1}{3}\cos^2{\phi_3\over 2}
  \cos^2{\phi_2\over 2}\sin{\phi_1}\sin{3t}\,\hat{\rho}(3\sigma)
  +\frac{1}{2}\cos^2{\phi_3\over 2}\sin{\phi_2}\cos{\phi_1}\sin{2t}
    \,\hat{\phi}(2\sigma-\theta_2)\]
\[\!\!\!\!\!\!\!\!\!\! +{1\over 4}\sin{\phi_3}\sin{\phi_2}
  \sin{\phi_1}\sin{2t}\,\hat{\rho}(2\sigma-\theta_3+\theta_2)
   -{1\over 2}\sin{\phi_3}\cos^2{\phi_2\over 2}\sin{\phi_1}\sin{2t}
    \sin{(2\sigma+\theta_3)}\hat z\]
\[\!\!\!\!\!\!\!\!\!\! -\cos^2{\phi_3\over 2}\sin^2{\phi_2\over 2}
  \sin{\phi_1}\sin{t}\,\hat{\rho}(\sigma-2\theta_2)
   -\sin^2{\phi_3\over 2}\sin^2{\phi_2\over 2}\sin{\phi_1}\sin{t}
    \,\hat{\rho}(\sigma-2\theta_3+2\theta_2)\]
\[\!\!\!\!\!\!\!\!\!\! +\sin{\phi_3}\cos{\phi_2}\cos{\phi_1}\sin{t}
  \,\hat{\phi}(\sigma-\theta_3)-\sin^2{\phi_3\over 2}
  \cos^2{\phi_2\over 2}\sin{\phi_1}\sin{t}\,\hat{\rho}
  (-\sigma-2\theta_3)\]
\[\!\!\!\!\!\!\!\!\!\! -\cos{\phi_3}\sin{\phi_2}\sin{\phi_1}\sin{t}
  \sin{(\sigma+\theta_2)}\hat z
   +\sin{\phi_3}\sin{\phi_2}\cos{\phi_1}\sin{t}\cos{(\sigma+
  \theta_3-\theta_2)}\hat z  \]
\[\!\!\!\!\!\!\!\!\!\! -{1\over 2}\sin{\phi_3}\sin{\phi_2}
  \sin{\phi_1}\, t\,\hat{\rho}(-\theta_3-\theta_2) -
  \sin^2{\phi_3\over 2}\sin{\phi_2}\cos{\phi_1}\, t\,\hat{\rho}
  (2\theta_3 -\theta_2)\]
\begin{equation} -[\cos{\phi_3}\cos{\phi_2}\cos{\phi_1} +\sin{\phi_3}
  \sin^2{\phi_2\over 2}
  \sin{\phi_1}\sin{(\theta_3 -2\theta_2)}]\, t\,\hat z \, ,
\end{equation}

or
\[\!\!\!\!\!\!\!\!\!\!\vec r=-{1\over 3}\cos^2{\phi_3\over 2}\cos^2
  {\phi_2\over 2}\sin{\phi_1}\cos{3\sigma}\,\hat{\phi}(-3t)
   -{1\over 2}\cos^2{\phi_3\over 2}\sin{\phi_2}\cos{\phi_1}\cos{2\sigma}
    \,\hat{\rho}(-2t-\theta_2) \]
\[\!\!\!\!\!\!\!\!\!\! +{1\over 4}\sin{\phi_3}\sin{\phi_2}
  \sin{\phi_1}\cos{2\sigma}\,\hat{\phi}(-2t-\theta_3+\theta_2)\]
\[\!\!\!\!\!\!\!\!\!\! -{1\over 2}\sin{\phi_3}\cos^2{\phi_2\over 2}\sin{\phi_1}
  \cos{2\sigma}\cos{(2t-\theta_3)} \hat z\]
\[\!\!\!\!\!\!\!\!\!\! -\cos^2{\phi_3\over 2}\sin^2{\phi_2\over 2}
  \sin{\phi_1}\cos{\sigma}\,\hat{\phi}(-t-2\theta_2)
   -\sin^2{\phi_3\over 2}\sin^2{\phi_2\over 2}\sin{\phi_1}
  \cos{\sigma}\,\hat{\phi}(-t-2\theta_3+2\theta_2) \]
\[\!\!\!\!\!\!\!\!\!\! -\sin{\phi_3}\cos{\phi_2}\cos{\phi_1}
  \cos{\sigma}\,\hat{\rho}(-t-\theta_3)+\sin^2{\phi_3\over 2}
  \cos^2{\phi_2\over 2}\sin{\phi_1}\cos{\sigma}\,\hat{\phi}
  (t-2\theta_3) \]
\[\!\!\!\!\!\!\!\!\!\! -\cos{\phi_3}\sin{\phi_2}\sin{\phi_1}
  \cos{\sigma}\cos{(t-\theta_2)}\hat z+\sin{\phi_3}\sin{\phi_2}
  \cos{\phi_1}\cos{\sigma}\sin{(t-\theta_3+\theta_2)}\hat z \]
\[\!\!\!\!\!\!\!\!\!\! -{1\over 2}\sin{\phi_3}\sin{\phi_2}
  \sin{\phi_1}\, t\,\hat{\rho}(-\theta_3-\theta_2) -
  \sin^2{\phi_3\over 2}\sin{\phi_2}\cos{\phi_1}\, t\,\hat{\rho}
  (2\theta_3 -\theta_2)\]
\begin{equation} -[\cos{\phi_3}\cos{\phi_2}\cos{\phi_1} +\sin{\phi_3}
  \sin^2{\phi_2\over 2}
     \sin{\phi_1}\sin{(\theta_3 -2\theta_2)}]\, t\,\hat z .
\end{equation}

\vskip2ex
\subsubsection{c.m. string}
\label{subsubsec:N=3CM}

There are many ways to set both zero harmonics equal to zero.
Because any combination of these string halves is possible, we
present only the equation for $\vec a$.

There are three solutions which produce a string with first,
second, and third harmonics.
\begin{equation} \phi_3=0, \ \phi_2={\pi\over 2},{3\pi\over 2} \, ,
\end{equation}
\[\vec a= [{1\over 6}\sin{\phi_1}\sin{3u}
         -(\pm_2){1\over 2}\cos{\phi_1}\cos{(2u-\theta_2)}
         +{1\over 2}\sin{\phi_1}\sin{(u-2\theta_2)}]\,\hat x \]
\[\quad   +[-{1\over 6}\sin{\phi_1}\cos{3u}
          -(\pm_2){1\over 2}\cos{\phi_1}\sin{(2u-\theta_2)}
          -{1\over 2}\sin{\phi_1}\cos{(u-2\theta_2)}]\,\hat y \]
\begin{equation} \quad   +[-(\pm_2)\sin{\phi_1}\cos{(u+\theta_2)} ]
  \,\hat z \, ,
\end{equation}

\begin{equation} \phi_3={\pi\over 2},{3\pi\over 2}, \ \phi_2=0 ,
\end{equation}
\[\vec a=[{1\over 6}\sin{\phi_1}\sin{3u}
       -(\pm_3)\cos{\phi_1}\cos{(u-\theta_3)}
       +{1\over 2}\sin{\phi_1}(\sin{(u+2\theta_3)}]\,\hat x \]
\[\quad  [-{1\over 6}\sin{\phi_1}\cos{3u}
          -(\pm_3)\cos{\phi_1}\sin{(u-\theta_3)}
          +{1\over 2}\sin{\phi_1}\cos{(u+2\theta_3)}]\,\hat y \]
\begin{equation} \quad  [-(\pm_3){1\over 2}\sin{\phi_1}\cos{(2u+
  \theta_3)}]\,\hat z \, ,
\end{equation}

\begin{equation}\phi_2=0, \ \phi_1={\pi\over 2},{3\pi\over 2},
\end{equation}
\[\vec a=(\pm_1)[{1\over 3}\cos^2{\phi_3\over 2}\sin{3u}
  +\sin^2{\phi_3\over 2}\sin{(u+2\theta_3)}] \,\hat x \]
\[\quad +(\pm_1)[-{1\over 3}\cos^2{\phi_3\over 2}\cos{3u}
  +\sin^2{\phi_3\over 2}\cos{(u+2\theta_3)}]\,\hat y \]
\begin{equation} \quad +(\pm_1)[-{1\over 2}\sin{\phi_3}
  \cos{(2u+\theta_3)}] \,\hat z \, ,
\end{equation}

The angles
\begin{equation} \phi_3=0, \ \phi_1={\pi\over 2},{3\pi\over 2},
\end{equation}
give a string with only first and third harmonics
\[\vec a=(\pm_1)[{1\over 3}\cos^2{\phi_2\over 2}\sin{3u}
          +\sin^2{\phi_2\over 2}\sin{(u-2\theta_2)}]\,\hat x\]
\[\quad +(\pm_1)[-{1\over 3}\cos^2{\phi_2\over 2}\cos{3u}
   -\sin^2{\phi_2\over 2}\cos{(u-2\theta_2)}]\,\hat y \]
\begin{equation}\quad+(\pm_1)[-\sin{\phi_2}\cos{(u+\theta_2)}]
  \,\hat z \, .
\end{equation}

The eight other solutions found reduce to strings with only a
first harmonic.  These solutions are
\begin{equation}
  \begin{array}{lll} \phi_3=\pi, & \phi_1={\pi\over 2},{3\pi\over 2},
  & \\ \phi_3={\pi\over 2},{3\pi\over 2}, & \phi_2=\pi, &
  \phi_1=0,\pi, \\
  \phi_3={\pi\over 2},{3\pi\over 2}, & \phi_2=\pi, & \theta_3=2
  \theta_2, \\ \phi_3={\pi\over 2},{3\pi\over 2}, & \phi_2=\pi, &
  \theta_3=2\theta_2+\pi, \\ \phi_2=\pi, & \phi_1={\pi\over 2},
  {3\pi\over 2}, & \theta_3=2\theta_2, \\ \phi_2=\pi, & \phi_1=
  {\pi\over 2},{3\pi\over 2}, & \theta_3=2\theta_2+\pi, \\
   \phi_2=\pi, & \theta_3=2\theta_2+{\pi\over 2}, & \phi_3=-\phi_1
  +{\pi\over 2},-\phi_1+{3\pi\over 2}, \\ \phi_2=\pi, & \theta_3=2
  \theta_2+{3\pi\over 2}, & \phi_3=\phi_1+{\pi\over 2},\phi_1+
  {3\pi\over 2}.
\label{eq:N=3CMCond}\end{array}\end{equation}

   An example of an N=3 string is given here,
$$\vec r={1\over 12}\sin{\phi_1}\left(\begin{array}{c}\cos{\alpha}\\
			\sin{\alpha}\\0\end{array}\right)\sin{3u}
	 -{1\over 12}\sin{\phi_1}\left(\begin{array}{c}
		-\sin{\alpha}\cos{\beta}\\ \cos{\alpha}\cos{\beta}\\
		\sin{\beta}\end{array}\right)\cos{3u}$$
$$\qquad +{1\over 4}\cos{\phi_1}\left(\begin{array}{c}
		\sin{\alpha}\cos{\beta}\\ -\cos{\alpha}\cos{\beta}\\
		-\sin{\beta}\end{array}\right)\sin{(2u-\theta_2)}
	 -{1\over 4}\cos{\phi_1}\left(\begin{array}{c}
		\cos{\alpha}\\ \sin{\alpha}\\ 0 \end{array}\right)
		\cos{(2u-\theta_2)}$$
$$\qquad +{1\over 4}\sin{\phi_1}\left(\begin{array}{c}\cos{\alpha}\\
		\sin{\alpha}\\ 0 \end{array}\right)\sin{(u-2\theta_2)}
	 -{1\over 4}\sin{\phi_1}\left(\begin{array}{c}
		-\sin{\alpha}\cos{\beta}\\ \cos{\alpha}\sin{\beta}\\
		\sin{\beta}\end{array}\right)\cos{(u-2\theta_2)}$$
$$\qquad \!\!\!\!\!\!\!\!\!\!\!\!\!\!\!\!\!\!\!\!\!\!\!\!\!\!\!\!\!\!
\!\!
         -{1\over 2}\sin{\phi_1}\left(\begin{array}{c}
		\sin{\alpha}\sin{\beta}\\ -\cos{\alpha}\sin{\beta}\\
		-\cos{\beta}\end{array}\right)\cos{(u+\theta_2)}
	 +{1\over 12}\left(\begin{array}{c}\sin{\phi'_1}\\ 0\\ 0
		\end{array}\right)\sin{3v}$$
$$\qquad -{1\over 12}\left(\begin{array}{c}0\\ \sin{\phi'_1}\\ 0
		\end{array}\right)\cos{3v}
	 +{1\over 4}\left(\begin{array}{c} 0\\ -\cos{\phi'_1}\\ 0
		\end{array}\right)\sin{(2v-\theta'_2)}$$
$$\qquad\ \ \ \ \
         -{1\over 4}\left(\begin{array}{c}\cos{\phi'_1}\\ 0\\ 0
		\end{array}\right)\cos{(2v-\theta'_2)}
	 +{1\over 4}\left(\begin{array}{c}\sin{\phi'_1}\\ 0\\ 0
	 	\end{array}\right)\sin{(v-\theta'_2)}$$
\begin{equation}
\qquad\ \ \ \ \ \ \ \ \ \
         -{1\over 4}\left(\begin{array}{c}0\\ \sin{\phi'_1}\\ 0
	 	\end{array}\right)\cos{(v-2\theta'_2)}
	 -{1\over 2}\left(\begin{array}{c}0\\ 0\\ \sin{\phi'_1}
		\end{array}\right)\cos{(v+\theta'_2)}.
\label{eq:N=3demo}\end{equation}
This string is displayed in Fig. 2 for a given choice of parameters.

\vskip5ex

\section{Previous Strings}
\label{sec:PrevStrings}

        In this section we rewrite other existing parametrizations
in terms of our rotation angles.  (These rewritten versions have
also been referenced earlier in~\cite{BD}.)
The Turok string~\cite{T}, shown in Fig. 3 for a particular choice of
parameters,
is a two-parameter string
which has a first and third harmonic in $\vec a$, and a first in
$\vec b$. Defining Turok's parameter $\alpha$ to be $\alpha \equiv
\sin^2{\eta\over 2}$, the string equation is
\[\vec r_T={1\over 2} \left( \begin{array}{c}
  {1\over 3}\sin^2{\eta\over 2}\sin{3u} +\cos^2{\eta\over 2}\sin{u}
  \\ -{1\over 3}\sin^2{\eta\over 2}\cos{3u} -\cos^2{\eta\over 2}
  \cos{u} \\ \sin{\eta}\cos{u}
  \end{array} \right)  \]
\begin{equation} \qquad\!\!  +{1\over 2} \left( \begin{array}{c}
   \sin{v} \\ -\cos{\phi}\cos{v} \\ -\sin{\phi}\cos{v}
   \end{array} \right).
\end{equation}
The product of rotations which yields the 1-3 half of this string
is
\begin{equation} \vec a^{\,\prime}_T= R_z(2u)R_x(\pi-\eta)R_z(u)
  \hat x,
\label{eq:Turokrot}\end{equation}
which corresponds to our complete product representation for $N=3$
with $\phi_3=0$, $\phi_2=\pi -\eta$, $\theta_2=0$,
$\phi_1={\pi\over 2}$, $\theta_1=-{\pi\over 2}$, using the
identity $R_z({\pi\over 2})R_x({\pi\over 2})R_z(-{\pi\over
2})\hat z=\hat x$.

The second string half is given by
\begin{equation} \vec b^{\,\prime}_T=R_x(\phi)R_z(v)\hat x.
\label{eq:Turokrot2}\end{equation}
The leftmost rotation is due to the fact that the Turok string is
not in what we call standard form. We see that this reduces to the
Kibble-Turok~\cite{KT} formula for $\phi=0$.

The Chen, DiCarlo, Hotes string \cite{CDH}, shown in Fig. 4 for
a given choice of paramterers, is a
1-3/1 string with an added parameter in $\vec a$, which contains the
Turok string as a limit.  We redefine the original CDH parameter
$\eta_{CDH}\equiv {\pi\over 2}-\eta$ to be consistent with our
definition of $\alpha$ in the Turok string.  The string equation
becomes
\[\vec r_{CDH}=\quad {1\over 6} \left( \begin{array}{c}
                 C\cos{\theta} \\ S \\ -C\sin{\theta}
                          \end{array} \right) \sin{3u}
          -{1\over 6} \left( \begin{array}{c}
                -S\cos{\theta} \\ C \\ S\sin{\theta}
                          \end{array} \right) \cos{3u} \]
\[\qquad  +{1\over 4} \left( \begin{array}{c}
                2-C\cos{\theta} \\ -S \\ C\sin{\theta}
                      \end{array} \right) \sin{u}
          -{1\over 4} \left( \begin{array}{c}
                3S\cos{\theta} \\ \cos{\theta}+\cos{\eta} \\
                 -\sin{\eta}(1-3\cos^2{\theta}) \end{array} \right)
                  \cos{u} \]
\begin{equation} \qquad  +{1\over 2} \left( \begin{array}{c}
                     1 \\ 0 \\ 0 \end{array} \right) \sin{v}
          -{1\over 2} \left( \begin{array}{c}
                     0 \\ \cos{\phi} \\ \sin{\phi}
                     \end{array} \right) \cos{v},  \end{equation}
where
\begin{equation} C=\cos{\theta}-\cos{\eta},
  \quad S=\sin{\theta}\sin{\eta}.
\end{equation}

The product representation is
\begin{equation} \vec a^{\,\prime}_{CDH}=R_y(\theta)R_z(\chi)R_z(2u)
  R_z(-\theta_2)R_x(\phi_2)R_z(\theta_2)R_z(u)\hat x,  \end{equation}
\begin{equation} \vec b^{\,\prime}_{CDH}=R_x(\phi)R_z(v)\hat x,
\end{equation}
where
\begin{equation} \cos{\chi}=C/(1-\cos{\theta}\cos{\eta}),
  \quad \sin{\chi}=S/(1-\cos{\theta}\cos{\eta}), \end{equation}
\begin{equation}\cos{\theta_2}=\cos{\theta}\sin{\eta}/d,
  \quad \sin{\theta_2}=\sin{\theta}/d, \end{equation}
\begin{equation} \cos{\phi_2}=-\cos{\theta}\cos{\eta},
  \quad \sin{\phi_2}=d, \end{equation}
\begin{equation} d=\pm(1-\cos^2{\theta}\cos^2{\eta})^{1\over 2}.
\end{equation}

DeLaney, Engle, and Scheick have derived a general five-parameter
1-3/1-3 string equation \cite{DES}
$$\vec r_{DES}={1\over 6} \left( \begin{array}{c}
  \sin{\phi} \\ -\cos{\phi}\cos{\gamma} \\ \cos{\phi}\sin{\gamma}
  \end{array} \right) \sin{3u}
 -{1\over 6} \left( \begin{array}{c}  \sin{\theta}\cos{\phi} \\
  \sin{\theta}\sin{\phi}\cos{\gamma}+\cos{\theta}\sin{\gamma} \\
 -\sin{\theta}\sin{\phi}\sin{\gamma}+\cos{\theta}\cos{\gamma}
\end{array} \right) \cos{3u} $$
\[\qquad\!\!\!\!\!  +{1\over 2} \left( \begin{array}{c}
  1-c\sin{\phi} \\ c\cos{\phi}\cos{\gamma} \\
  -c\cos{\phi}\sin{\gamma} \end{array} \right) \sin{u}
  -{1\over 2} \left( \begin{array}{c}
  x \\ y\cos{\gamma} +z\sin{\gamma} \\ -y\sin{\gamma} +z\cos{\gamma}
          \end{array} \right) \cos{u} \]
\[\qquad\!\!\!\!\!  +{1\over 6} \left( \begin{array}{c}
  \sin{\phi'} \\ -\cos{\phi'} \\ 0 \end{array} \right) \sin{3v}
 -{1\over 6} \left( \begin{array}{c}
  \sin{\theta'}\cos{\phi'} \\ \sin{\theta'}\sin{\phi'} \\
  \cos{\theta'}  \end{array} \right) \cos{3v} \]
\begin{equation} \qquad\!\!\!\!\!  +{1\over 2} \left(
  \begin{array}{c} 1-c'\sin{\phi'} \\ c'\cos{\phi'} \\ 0
  \end{array} \right) \sin{v}
 -{1\over 2} \left( \begin{array}{c}
    x' \\ y' \\ z'
  \end{array} \right) \cos{v},
\end{equation}
with
\[ x=-3c\sin{\theta}\cos{\phi},\]
\[ y=\sin{\theta}(1-3c\sin{\phi}),\]
\[ z=\cos{\theta}\sin{\phi}-c\sec{\theta}(1-3\sin^2{\theta}),\]
\begin{equation} c=\cos^2{\theta}(1+\sin{\phi})/2,  \end{equation}
and with the same relationships between the primed variables.

In terms of the product representation,
\begin{equation}
  \vec a^{\,\prime}_{DES}=R_x(-\gamma)R_z(\phi-{\pi\over 2})
  R_x({\pi\over 2}-\theta)R_z(2u)R_z(-\theta_2)R_x(\phi_2)
  R_z(\theta_2)R_z(u) \hat x,
\end{equation}
\begin{equation} \vec b^{\,\prime}_{DES}=R_z(\phi'-{\pi\over 2})
  R_x({\pi\over 2}-\theta')R_z(2v)R_z(-\theta'_2)R_x(\phi'_2)
  R_z(\theta'_2)R_z(v)\hat x,
\end{equation}
with
\begin{equation} \cos{\theta_2}=\pm\sin{\theta}(1+\sin{\phi})/f,
   \quad \sin{\theta_2}=\mp\cos{\phi}/f, \end{equation}
\begin{equation} \cos{\phi_2}=\cos^2{\theta}\sin{\phi}
  -\sin^2{\theta}, \quad \sin{\phi_2}=\pm f\cos{\theta},
\end{equation}
\begin{equation}
  f=[\cos^2{\phi}+\sin^2{\theta}(1+\sin{\phi})^2]^{1\over 2},
\end{equation}
and similarly for the primed angles. This string is displayed in Fig.
5 for a given choice of parameters.

These early parametrizations contained only odd harmonics.  For this
reason, they satisfy the symmetry relation $\vec{r}(\sigma+\pi,t)
= -\vec{r}(\sigma,t)$.  On the other hand,
in Ref.\cite{Bu} Burden introduced a simple class of
strings with m and n harmonics in the left and right sectors, respectively,
\begin{equation} \begin{array}{l}
\vec{a}\,'(u) = cos(mu)\hat{x} - sin(mu)\hat{z}, \\
\vec{b}\,'(v) = cos(nv)(cos\Psi \hat{x}+sin\Psi \hat{y})
	-sin(nv)\hat{z}.
\end{array} \end{equation}
For $m$ and $n$ relatively prime, this symmetry no longer holds, and
gravitational radiation is no longer suppressed \cite{Bu}.
We can describe this class
of strings as a subclass of our general parametrization in the following
way,
\begin{equation} \begin{array}{l}
\vec{a}\,'(u) = [R_{y}(u)]^{m}\,\hat{x}, \\
\vec{b}\,'(v) = R_{z}(\Psi)[R_{y}(v)]^{n}\,\hat{x}.
\end{array} \end{equation}

Another string without the aforementioned symmetry is
the Vachaspati-Vilenkin string~\cite{VV}, an example of which is shown in
Fig. 6.  This has first,
second, and third harmonics in the left-going string half, and a
first in the right-going half:
\begin{equation} \vec r_{VV} = {1\over 2} \left( \begin{array}{c}
   -{1\over 3}\alpha\sin{3u} +(1-\alpha)\sin{u} +\sin{v} \\
   -{1\over 3}\alpha\cos{3u} -(1-\alpha)\cos{u} -\cos{\phi}\cos{v} \\
   \sqrt{\alpha (1-\alpha)}\,\sin{2u} -\sin{\phi}\cos{v}
  \end{array} \right),  \end{equation}
where $0\leq\alpha\leq 1$, and $-\pi\leq\phi\leq\pi$.  Defining
$\phi_3$ by the relation $\alpha\equiv\cos^2{\phi_3\over 2}$, the
product representation gives
\begin{equation} \vec a_{VV}^{\,\prime}=-R_x(\pi)R_z(u)
  R_z(-{\pi\over 2})R_x(\phi_3)R_z({\pi\over 2})R_z(2u)\hat x,
\end{equation}
\begin{equation} \vec b_{VV}^{\,\prime}=R_x(\phi)R_z(v)\hat x.
\end{equation}

Kibble, Garfinkle and Vachaspati have derived a
formula for cuspless loops~\cite{GV}, shown in Fig. 7, again with arbitrary
parameters:
\[\vec r_{GV} = {1\over 2}{1\over p^2+2}{1\over 2p^2+1} \left(
  \begin{array}{c}{p^2\over 4}\sin{4u} +(p^2+1)^2\sin{2u} \\
  -{2\sqrt{2}\over 3}p\cos{3u} -2\sqrt{2}p(p^2+2)\cos{u} \\
  -{p^2\over 4}\cos{4u} -((p^2+1)^2-2)\cos{2u}
  \end{array} \right)\]
\begin{equation} \ \ \ \ \ \ \ +{1\over 2}{1\over p^2+2}
  {1\over 2p^2+1} \left( \begin{array}{c}
   {p^2\over 4}\sin{4v} +(p^2+1)^2\sin{2v}\\
   {p^2\over 4}\cos{4v} +((p^2+1)^2-2)\cos{2v}\\
   {2\sqrt{2}\over 3}p\cos{3v} +2\sqrt{2}p(p^2+2)\cos{v}
  \end{array} \right),  \end{equation}
where $p$ is a constant.
In terms of the product representation, this is
\begin{equation} \vec a^{\,\prime}=R_x({\pi\over 2})R_z(u)
  R_x(\phi_4)R_z(2u)R_x(\phi_2)R_z(u)\hat x,
\end{equation}
\begin{equation} \vec b^{\,\prime}=-R_y(\pi)R_z(v)R_x(\phi_4)
  R_z(2v)R_x(\phi_2)R_z(v)\hat x,
\end{equation}
where $\phi_4$ and $\phi_2$ are related to $p$ by the definition
\begin{equation} p\equiv -\sqrt{2}\cot{\phi_4\over 2}\equiv
  \tan{\phi_2\over 2}/\sqrt{2},
\end{equation}
so that
\begin{equation} \sin{\phi_4}=-2\sqrt{2}p/(p^2+2), \quad
  \cos{\phi_4}=(p^2-2)/(p^2+2),
\end{equation}
\begin{equation} \sin{\phi_2}=2\sqrt{2}p/(2p^2+1), \quad
  \cos{\phi_2}=-(2p^2-1)/(2p^2+1).
\end{equation}

\vskip5ex

\section{Kink Parametrizations}

While a cusp is a point on a string where $\vec{r}_{\sigma}=0$, a kink
is a discontinuity in $\vec{r}_{\sigma}$.  A discontinuity of this
type may occur after intercommutation (e.g., if a loop crosses itself,
it splits and reconnects as two closed loops).  The produced kink splits
into right and left traveling kinks residing in $\vec{a}$
and $\vec{b}$, respectively.  We should like to define these as right kinks
and left kinks, for short.  Indeed, a picture of a loop with both a left and
right kink will show only one bend at the instant when the two pass through
each other.

\subsection{Paired Kinks}

	Parametrization of loops with kinks has been investigated by
Garfinkle and Vachaspati \cite{GV}.  Using a greatest integer function,
they were able to put symmetric sets of kinks (i.e., two symmetric pairs
of left and right kinks) in an extension of Burden trajectories \cite{Bu}.
With four parameters, $p,q,\delta$ and $\Psi$, define
\begin{equation}
\begin{array}{l}
\alpha = \pi (1-p)[2(\sigma-t)/L], \\
\beta = \pi (1-q)[2(\sigma+t)/L] + \delta, \end{array} \end{equation}
where $[x]$ is the greatest integer less than or equal to x.  The
Garfinkle-Vachaspati solution is given by
\begin{equation}
\begin{array}{l}
\vec{a}\,' = sin(2\pi p(\sigma-t)/L + \alpha)\hat{x}
	   + cos(2\pi p(\sigma-t)/L + \alpha)\hat{z}, \\
\vec{b}\,' = sin(2\pi q(\sigma+t)/L + \beta)[cos\Psi \hat{x}
	   +sin\Psi \hat{y}] + cos(2\pi q(\sigma+t)/L + \beta)\hat{z}.
\end{array} \end{equation}

\subsection{Symmetric Kinks}

The paired kinks can be generalized to a symmetric set.  Define
\begin{equation}
f(u) = p\,u + \frac{2\pi}{n}(1-p)[u/\frac{2\pi}{n}]
\end{equation}
where $p \in (0,1)$ is real and $n > 1$ is an integer.  $[x]$ is defined
to be the greatest integer less than or equal to $x$.  Then a string with
n symmetric kinks $(n > 1)$ is given by
\begin{equation}
\vec{a}' = (cos(f(u)),sin(f(u)),0).
\end{equation}
Due to the symmetric placement of the kinks, the integral of $\vec{a}'$
over $u = 0$ to $2\pi$ clearly vanishes.

For an even number of symmetric kinks (n = even), this parametrization
can be generalized in the manner discussed in section III.  For example,
a large multiparameter set of symmetric kinked strings is given by
\begin{equation}
\vec{a}' = \left( \prod_{i}R_{x_{i}}(2m_{i}u + \beta_{i})\right)
           \, R_{x_{3}}(f(u)) \, \hat{x_{1}},
\end{equation}
where the $x_{i}, m_{i}$, and $\beta_{i}$ are, respectively, arbitrary axes,
integers, and angles.

\subsection{Single Kinks}
 Parametrizations having a single kink in either $\vec{a}$ or $\vec{b}$, (left
or right kink), or both, instead of a pair of symmetric kinks such as shown
above
, are difficult to express in closed form.  To see this, consider the
continuity constraint,
\begin{equation}
\int_{0}^{2\pi} \vec{r}_{\sigma}(\sigma) d\sigma = 0.
\end{equation}
For symmetric kinks, this constraint becomes trivial as the first
half of the integral cancels the second half for all components.

	For the single kink string we can satisfy this continuity constraint in the
following way.  As before, we
split $\vec{r}$ into its left-going and right-going
modes.  Letting $\vec{b}$ be smooth, we derive the
discontinuity solely from $\vec{a}\,'$.  We ask that $\vec{a}\,'$ lie in
the x-y plane and that the discontinuity occur at u = $2n\pi$ where
n is an integer.  Letting
\begin{equation}
\vec{a}\,'(0) = \left( \begin{array}{c} cos\,\alpha \\ sin\,\alpha \\ 0
\end{array} \right) \; \; ; \; \; \vec{a}\,'(2\pi) = \left( \begin{array}{c}
cos\,\alpha \\ -sin\,\alpha \\ 0 \end{array} \right) , \end{equation}
will give us a kink with a discontinuity angle varying
with time, and depending on $\vec{b}\,'$ and an arbitrary constant $\alpha$.
The problem now
lies in finding a parametrization of the path $\vec{a}\,'$ takes along the unit
circle such that its integral vanishes.

	We can set
\begin{equation} \vec{a}\,'(u) = \left( \begin{array}{c} cos(\alpha +
\frac{\pi-\alpha}{\pi}f(u)) \\ sin(\alpha+\frac{\pi-\alpha}{\pi}f(u)) \\
0 \end{array} \right), \end{equation}
asking that $f(0) = 0$ and $f(2\pi) = 2\pi$.  To get the integral to vanish,
one component can be taken care of by symmetry condition on $f$.  For the
other component, we need $f$ to change slowly over a smaller part of the
circle and then quickly over the rest.   A simple example is
\begin{equation}
f(u) = u - \delta sin\,u\,.
\end{equation}
$\delta$ is positive and constant.  The integration constraint results in a
nontrivial relation between
$\alpha$ and $\delta$,
\begin{equation}
tan\,\alpha = \frac{J_{\frac{\pi-\alpha}{\pi}}(-\delta\frac{\pi-\alpha}{\pi})
	-cos \alpha \,J_{\frac{\pi-\alpha}{\pi}}(\delta\frac{\pi-\alpha}{\pi})
	-sin \alpha \,E_{\frac{\pi-\alpha}{\pi}}(\delta\frac{\pi-\alpha}{\pi})}
	{E_{\frac{\pi-\alpha}{\pi}}(-\delta\frac{\pi-\alpha}{\pi})
	+sin \alpha \,J_{\frac{\pi-\alpha}{\pi}}(\delta\frac{\pi-\alpha}{\pi})
	-cos \alpha \,E_{\frac{\pi-\alpha}{\pi}}(\delta\frac{\pi-\alpha}{\pi})},
\end{equation}
where the Anger function and Weber function, $J$ and $E$ respectively, are
given by
\begin{equation}
\begin{array}{l}
J_{\nu}(z) = \frac{1}{\pi}\int_{0}^{\pi} cos(\nu s - z sin s) ds, \\
E_{\nu}(z) = \frac{1}{\pi}\int_{0}^{\pi} cos(\nu s - z sin s) ds.
\end{array} \end{equation}

This transcendental relation is well defined, and has as a sample numerical
solution, $\delta$ = 0.736 for $\alpha = \frac{\pi }{4}$.  This solution
generalizes

easily to an infinite parameter set of string solutions, where $f(u)$ is
\begin{equation}
f(u) = u + \sum_{n=0}^{\infty} \delta_{n}sin(n u),
\end{equation}
subject to a single integral condition as a generalization of equation (97).

Other modulating functions can also be considered, such as polynomial functions
\begin{equation}
f(u) = u + \sum_{n = odd} \alpha_{n}\left[ \frac{(u-\pi )}{\pi }\right]^{n},
\end{equation}
subject to two conditions.  The first being the continuity condition, which is
transcendental in the arguments $\alpha_{n}$, and the condition that
$f(0) = 0$, which means that the sum of the
coefficients $\alpha_{n}$ vanishes.

\section{Kinks After Intercommutation}
In the previous section we have discussed representations of strings with
single kinks traveling in one direction around the string loop, where only
$\vec{a}\,'
$ or $\vec{b}\,'$
but not both, contains a discontinuity.  We have also mentioned, however, that
this description is not adequate for describing a string kinked as a result of
intercommutation.  We will first consider a generic case of intercommutation.
This is followed by an explicit analytic example.

\subsection{A construction algorithm}
At the point of intercommutation (say, $\sigma = t=0$), locally the two legs of
the newly formed kink describe a plane in 3-space.  The directions of these
legs
 are given by $\vec{r}_{\sigma}$ to the left
and right of the kink, and we define these vectors as $\vec{r}_{\sigma}
(-\epsilon ,0 )$ and $\vec{r}_{\sigma}(+\epsilon ,0 )$, respectively.  Because
the two legs of the kink were originally parts of string segments crossing
through each other, each leg moves transverse to itself and the plane, and
opposite,
generally, to each other.  That is, the
initial conditions require that $\vec{r}_{t}$ has a non-zero
component transverse to this plane that changes in sign when crossing
through the kink.  We will show that this initial condition is
contradictory to the conditions of a string containing only a single left
and right kink.

	Consider the initial
time-derivatives of $\vec{r}$ to the left and right of the kink,
$\vec{r}_{t}(-\epsilon ,0 )$ and $\vec{r}_{t}(+\epsilon ,0 )$, respectively.
Suppose only $\vec{a}\,'$ is discontinuous, as in the examples in the previous
section.  From eq.(1) we have,
\begin{equation} \begin{array}{l}
\vec{r}_{\sigma}(-\epsilon ,0 ) = \frac{1}{2} [\vec{a}\,'(-\epsilon )
			+ \vec{b}\,'(0)] , \\
\vec{r}_{\sigma}(+\epsilon ,0 ) = \frac{1}{2} [\vec{a}\,'(+\epsilon )
			+ \vec{b}\,'(0)] , \\
\vec{r}_{t}(-\epsilon ,0 ) = \frac{1}{2} [- \vec{a}\,'(-\epsilon )
			+ \vec{b}\,'(0)] , \\
\vec{r}_{t}(+\epsilon ,0 ) = \frac{1}{2} [- \vec{a}\,'(+\epsilon )
			+ \vec{b}\,'(0)] .
\end{array} \end{equation}
Noticing that the right-hand-sides contain only 3 independent vectors, we can
write one in terms of the others,
\begin{equation}
\vec{r}_{t}(+\epsilon ,0 ) = \vec{r}_{t}(-\epsilon ,0 ) + \vec{r}_{\sigma}
	(+\epsilon ,0 ) - \vec{r}_{\sigma}(-\epsilon ,0 ).
\end{equation}
And clearly since the last two vectors on the right-hand-side of eq.(105) lie
in the above-mentioned
plane, the component of $\vec{r}_{t}$ transverse to this plane must be
equal on both sides of the kink, in contradiction with the initial motion.  We
see therefore that discontinuities must occur in both $\vec{a}\,'$ and
$\vec{b}\,'$ to correctly describe a kink after intercommutation.

Under
these circumstances, the kink will split into left and right moving parts
(left and right kinks)
after intercommutation.  Given a string parametrization of a string that
self-intersects, we should like to show how one constructs the parametrization
for the
daughter loops.  Consider a low-harmonic
string that at some point in time crosses itself.  Let this crossing occur
at $t = \tau$, for $\sigma = \sigma_{0}, \sigma_{1}$.  To describe the
resulting
motion, we concentrate on one of the daughter loops.  Note that at
the point of intercommutation, this loop is described by its left and
right moving parts:  $\vec{a}(u), u \in [\sigma_{0}-\tau,\sigma_{1}-\tau]$
and $\vec{b}(v), v \in [\sigma_{0}+\tau,\sigma_{1}+\tau]$ , respectively.  We
can define a new string with $\vec{A},\, \vec{B}$ , of invariant length
$\Delta\sigma = \sigma_{1}-\sigma_{0}$,  in terms
of these functions $\vec{a}, \vec{b}$.  We ask that the left and
right moving parts of this new string, $\vec{A},\, \vec{B}$,  take the same
values as $\vec{a}$
and $\vec{b}$ over the intervals given above, respectively, and demand that
their derivatives $\vec{A}\,',\, \vec{B}\,'$
be periodic with period $\Delta\sigma$.

$\vec{A},\, \vec{B}$
will not generally be periodic, with the daughter usually
acquiring some center of mass velocity.  To wit, let
\begin{equation}
\vec{k} = \vec{a}(\sigma_{1}-\tau)-\vec{a}(\sigma_{0}-\tau)
	 = -( \vec{b}(\sigma_{1}+\tau)-\vec{b}(\sigma_{0}+\tau) ).
\end{equation}
$\vec{k}$ determines the resulting c.m. velocity of the piece.
Define $\vec{A},\, \vec{B}$ in the following way.  For $s \in
[0,\Delta\sigma]$, set
\begin{equation} \begin{array}{l}
\vec{A}(s) = \vec{a}(\sigma_{0}-\tau+s),\\
\vec{B}(s) = \vec{b}(\sigma_{0}+\tau+s).
\end{array} \end{equation}
For $s = n\Delta\sigma + s', s' \in [0,\Delta\sigma], n \in {\bf Z}$
we let
\begin{equation} \begin{array}{l}
\vec{A}(s) = \vec{a}(\sigma_{0}-\tau+s) + n\vec{k}, \\
\vec{B}(s) = \vec{b}(\sigma_{0}+\tau+s) - n\vec{k}.
\end{array} \end{equation}

The functions $\vec{A}$ and $\vec{B}$ in eq.(108)
describe a closed string of
period $\Delta\sigma$ containing a single kink at the time
of intercommutation that subsequently splits into left and right
moving kinks.  The closed-loop trajectory
\begin{equation}
\vec{r}(\sigma,t) =
	\frac{1}{2}(\vec{A}(\sigma-t)+\vec{B}(\sigma+t))
\end{equation}
satisfies the gauge conditions discussed in Section II.  The loop moves
with c.m. velocity $-\vec{k}/\Delta\sigma$, corresponding to the shift
relation
 \begin{equation}
\vec{r}(\sigma,t+\Delta\sigma) =
	\vec{r}(\sigma,t) - \vec{k} .
\end{equation}

We thus have a procedure for determining the equation of motion for a
cosmic string after intercommutation.  Given the harmonic parameterization
for a string, one first needs to calculate the points of self-intersection
(see Ref. \cite{CDH,DES,Em1} for typical calculation).  Then the above
procedure can be used to find an analytical expression for the resulting
string motion.

\subsection{Example}
	The construction of a loop with a single kink has been
discussed in the previous section.  It was explained how a kink could
be described in terms of a phase modification of a low harmonic form.  Here
we wish to look at the more realistic situation where a daughter loop is
produced upon the self-intersection of a closed low-harmonic string.

	In Reference \cite{CDH}, the range of parameters have been
carefully determined for which self-intersections occur in the Turok
string of eq.(62).  We shall follow the evolution of one of these
strings whose parameters lie in this range, before and after
self-intersection.  The harmonic forms can be adapted according to the
equations of the above subsection in order to describe the daughter loops.

	Using the techniques of Reference \cite{CDH}, we find, for our
example, that a Turok-string self-intersection occurs for $\alpha = 0.5,
\phi = 9\pi/20$ at $\tau = 4.88707, \sigma_{0} = -.404027, \sigma_{1} =
.333862$.  One can add $\pi$ to the $\sigma$ values to describe the other
self-intersection occurring  simultaneously (due to the  symmetry of this
class of loops).  This example will subsequently split into three subloops.
Using the aforementioned procedure, we have calculated the resulting
disintegration of the string.  Fig. 8 displays the parent string at the
point of self-intersection.  Fig. 9 shows the split, and Fig. 10 is a
close-up of the point of intercommutation after the separation.

	The evolution of the example is faithful to the gauge conditions
chosen (and described in Section II).  All strings segments have the
appropriate transverse motion and unit energy density, in view of
the fact that both $\vec{A}\,'$ and $\vec{B}\,'$ are unit vectors.
The two outside daughter loops have indeed acquired c.m. motion (with
one kink in each) but the middle daughter (with two symmetric kinks)
having none.  Of course, the number of left and right kinks is twice the number
of kinks.  Correspondingly the middle loop spins after the split,
conserving angular momentum.  The outer loops are more circular and
have smaller periods; they are observed to shrink rapidly down to rather
small size, as seen in Fig. 9.  Finally, we note that as the kink separates
into the left and right kinks, the string segment between the
two kinks is curved.  In general, there will not be a straight line
between the left and right kinks.

\section{Conclusions}
\label{sec:Conclusion}

        The solutions to the classical relativistic string model
for a fixed number of low harmonics are useful in a number of
different cosmic string applications.  Starting with the study
of a simple first plus third harmonic string~\cite{KT}, the issue of
self-intersections has been addressed for increasingly more
general
parametrizations~\cite{T,CT,CDH,Th,Al,Ho,Yo,DES,Em1}.

        Harmonic parametrizations have been called upon in
the calculation of string angular momentum~\cite{T,DES},
models of cosmic strings in an expanding universe~\cite{T2,T3}
and also of the string gravitational radiation with and without
kinks~\cite{VV2,Va,Du,Cre,AS}.
Gravitational particle production of
cosmic strings
has been studied using harmonic forms \cite{Ga}.
It has been observed that string trajectories containing only odd harmonics
do not emit gravitational radiation \cite{Bu,SQSP}.
It is hoped that the more general trajectories presented in this
paper will aid in giving a more realistic description of the gravitational
radiation of cosmic strings.
The electromagnetic self-interaction of a string has been calculated
using harmonic parametrizations~\cite{AVV,PSS,Am}.
The probability that harmonically parametrized
string loops will collapse to black holes
has also been addressed~\cite{PZ} and the interaction between harmonic
strings and domain walls has been studied \cite{Vile}. Solutions
incorporating these
parametrizations have the advantage that analytic precision is
not sacrificed for increasingly complex structure.

{\large\bf Acknowledgements}

We are grateful to Bruce Allen, Bob Scherrer and Alex Vilenkin for various
discussions about various topics related to this paper.  One of us (S.H.) would
like to thank Bob Hotes for the
warm accommodations provided him while working on this project.
This research was supported in part by the National Science Foundation.

\vfill
\newpage
{\large\centerline{\bf Figure Captions.}}
\vskip5ex
{\large\bf 1.} The N=1 C.M. string with $\phi = {\pi\over 2}$ at $t = 0$.

{\large\bf 2.} An N=3 C.M. string with $\alpha = 1.589$,
  $\beta = 1.712$, $\theta = 1.23$, $\theta^{\,\prime} = 3.011$,
  $\phi = 0.785$, and $\phi^{\,\prime} = 2.901$ at $t = 2$.

{\large\bf 3.} The Turok-PZ string with $\alpha=0.8$ and
  $\phi = {\pi\over 2}$ at $t = 0$.

{\large\bf 4.} The CDH string with $\theta = {{4\pi}\over 10}$,
  $\eta = 1.60160$, and $\phi = 2.64460$ at $t = 2.03$.

{\large\bf 5.} The DES string with $\theta = {{4\pi}\over 3}$,
  $\theta^{\,\prime} = 2.03$, $\phi = 0.64460$, $\phi^{\,\prime}
  = 2.2345$, and $\gamma = {{7\pi}\over 5}$ at $t = 2.0$.

{\large\bf 6.} The VV string with $\alpha = 0.6$ and
  $\phi = 2.64460$ at $t = 0$.

{\large\bf 7.} The GV cuspless, kinkless string with
  $p = 0.2$ at $t = {\pi\over 2}$.

%{\large\bf 8.} The GV kinky string with $p = 0.5$, $q = 0.5$,
%  and $\delta = 0$ at $t = {\pi\over 2}$.

{\large\bf 8.} Self-intersection of a Turok string.

{\large\bf 9.} The Turok string shortly after intercommutation.

{\large\bf 10.} A close-up of the intercommutation region.

\begin{thebibliography}{99}
\bibitem{Ki} T.W.B. Kibble, J. Phys.
  {\bf A9}, 1387 (1976).
\bibitem{COBE} See for example,
  D. Bennett, A. Stebbins, F. Bouchet, PRINT-92-0240, Jun 1992, 10pp.
  and L. Perivolaropoulos, Phys. Lett. {\bf B298}, 305 (1993).
\bibitem{Sh} E.P.S. Shellard, Nucl. Phys.
  {\bf B283}, 624 (1987).
\bibitem{Ma} R.A. Matzner, Comput. Phys.
  {\bf 2}, 51 (1988).
\bibitem{LM} P. Laguna, and R.A. Matzner, Phys. Rev. Lett.
  {\bf 62}, 1948 (1989).
\bibitem{GV} D. Garfinkle, and T. Vachaspati, Phys. Rev.
  {\bf D36}, 2229 (1987).
\bibitem{Qu} J.M. Quashnock, and D.N. Spergel, Phys. Rev.
  {\bf D42}, 2505 (1990).
\bibitem{KT} T.W.B. Kibble and N. Turok, Phys. Lett.
  {\bf 116B}, 141 (1982).
\bibitem{T} N. Turok, Nucl. Phys.
  {\bf B242}, 520 (1984).
\bibitem{Bu} C.J. Burden, Phys. Lett.
  {\bf B164}, 277 (1985).
\bibitem{VV} A. Vilenkin and T. Vachaspati, Phys. Rev. Lett.
  {\bf 58}, 1041 (1987).
\bibitem{CDH} A.L. Chen, D.A. DiCarlo, and S.A. Hotes, Phys. Rev.
  {\bf D37}, 863 (1988).
\bibitem{DES} D.B. Delaney, K.A. Engle, and X. Scheick, Phys. Rev.
  {\bf D41}, 1775 (1990).
\bibitem{Br} R.W. Brown, {\em The Formation and Evolution of
  Cosmic Strings}, edited by G. Gibbons, S. Hawking, and
  T. Vachaspati (Cambridge U. P., Cambridge, 1990), p. 127.
\bibitem{BD} R.W. Brown, and D.B. DeLaney, Phys. Rev. Lett.
  {\bf 63}, 474 (1989).
\bibitem{BCD} R.W. Brown, M.E. Convery, and D.B. DeLaney,
  J. Math. Phys. {\bf 32} (7), 1674 (1991).
\bibitem{BRT} R.W. Brown, E.M. Rains, and C.C. Taylor,
  Class. and Quant. Grav. {\bf 8}, 1245 (1991).
\bibitem{Sch} R.J. Scherrer and W.H. Press, Phys. Rev. {\bf D39}, 371 (1989).
\bibitem{All} B. Allen and R.R. Caldwell, Phys. Rev. {\bf D43}, 3173 (1991).
\bibitem{GGRT} P. Goddard, J. Goldstone, C. Rebbi, and C.B. Thorn,
  Nucl. Phys., {\bf B56}, 109 (1973).
\bibitem{CT} E.J. Copeland, and N. Turok, Phys. Lett.
  {\bf B172} (2), 129 (1986).
\bibitem{Th} C. Thompson, Phys. Rev. {\bf D37}, 283 (1988).
\bibitem{Al} A. Albrecht and T. York, Phys. Rev. {\bf D38}, 2958 (1988).
\bibitem{Ho} D. Hochberg, Nucl. Phys. {\bf B319}, 709 (1989).
\bibitem{Yo} T. York, Phys. Rev. {\bf D40}, 277 (1989).
\bibitem{Em1} F. Embacher, UWThPh-1992-14,15, March 26, 1992.
\bibitem{T2} N. Turok, Phys. Lett. {\bf B123}, 387 (1983).
\bibitem{T3} N. Turok and P. Bhattacharjee, Phys. Rev. {\bf D29},
  1557 (1984).
\bibitem{VV2} A. Vilenkin , and T. Vachaspati, Phys. Rev.
  {\bf D31}, 3052 (1985).
\bibitem{Va} T. Vachaspati, Phys. Rev. {\bf D39}, 1768 (1989).
\bibitem{Du} R. Durrer, Phys. Rev. {\bf B328}, 238 (1989) and PUPT-90-1165,
  (1990).
\bibitem{Cre} A. Cresswell and R.L. Zimmerman, Phys. Rev. {\bf D42}, 2527
   (1990).
\bibitem{AS} B. Allen and E.P.S. Shellard, Phys. Rev.
  {\bf D45}, 1898 (1992).
\bibitem{Ga} J. Garriga, D. Harari, and E. Verdaguer, Nucl. Phys.
  {\bf B339}, 560 (1990).
\bibitem{SQSP} R.J. Scherrer, J.M. Quashnock, D.N. Spergel and W.H. Press,
  Phys. Rev. {\bf D42}, 1908 (1990).
%\bibitem{St} A. Stebbins, Ap. J. {\bf 327}, 584 (1988).
\bibitem{AVV} M. Aryal, A. Vilenkin, and T. Vachaspati, Phys. Lett.
  {\bf B194}, 25 (1987).
\bibitem{PSS} W. Press, R. Scherrer, and D. Spergel, Phys. Rev.
  {\bf D39}, 379 (1989).
\bibitem{Am} P. Amsterdamski, Phys. Rev.
  {\bf D39}, 1524 (1989).
\bibitem{PZ} A. Polnarev and R. Zembowicz, Phys. Rev. Lett.
  {\bf 43}, 1106 (1991).
\bibitem{Vile} A. Vilenkin, Physics Reports {\bf 121}, 263 (1985).

\end{thebibliography}
\end{document}